\documentclass[conference]{IEEEtran}
\usepackage{cite}
\usepackage{comment}
\usepackage{amsmath,amssymb,amsfonts,amsthm}
\usepackage{algorithmic}
\usepackage{graphicx}
\usepackage[hang,flushmargin]{footmisc}
\usepackage{textcomp}
\usepackage{xcolor}
\usepackage{tikz}
\usetikzlibrary{shapes,arrows}
\usepackage[ruled,vlined]{algorithm2e}
\usepackage{booktabs}

\def\BibTeX{{\rm B\kern-.05em{\sc i\kern-.025em b}\kern-.08em
    T\kern-.1667em\lower.7ex\hbox{E}\kern-.125emX}}
\begin{document}

\title{Energy Efficient Communications in RIS-assisted UAV Networks Based on Genetic Algorithm \\
}

\author{\IEEEauthorblockN{Mohammad Javad-Kalbasi}
\IEEEauthorblockA{\textit{Department of Electrical Engineering} \\
\textit{University of Toronto}\\
Toronto, Canada \\
mohammad.javadkalbasi@mail.utoronto.ca}
\and
\IEEEauthorblockN{Mohammed S. Al-Abiad}
\IEEEauthorblockA{\textit{Department of Electrical Engineering} \\
\textit{University of Toronto}\\
Toronto, Canada \\
mohammed.saif@utoronto.ca}
\and
\IEEEauthorblockN{Shahrokh Valaee}
\IEEEauthorblockA{\textit{Department of Electrical Engineering} \\
\textit{University of Toronto}\\
Toronto, Canada \\
valaee@ece.utoronto.ca}}
 
\maketitle
\IEEEpubidadjcol
\begin{abstract}
This paper proposes a solution for energy-efficient communication in reconfigurable intelligent surface (RIS)-assisted unmanned aerial vehicle (UAV) networks. The limited battery life of UAVs is a major concern for their sustainable operation, and RIS has emerged as a promising solution to reducing the energy consumption of communication systems. The paper formulates the problem of maximizing the energy efficiency of the network as a mixed integer non-linear program, in which UAV placement, UAV beamforming, On-Off strategy of RIS elements, and phase shift of RIS elements are optimized. The proposed solution utilizes the block coordinate descent approach and a combination of continuous and binary genetic algorithms. Moreover, for optimizing the UAV placement, Adam optimizer is used. The simulation results show that the proposed solution outperforms the existing literature. Specifically, we compared the proposed method with the successive convex approximation (SCA) approach for optimizing the phase shift of RIS elements. 
\end{abstract}


\section{Introduction}
\footnote{This work was supported in part by funding from the Innovation for Defence Excellence and Security (IDEaS) program from the Department of National Defence (DND).}In recent years, unmanned aerial vehicles (UAVs) have gained immense popularity due to the high likelihood of establishing line-of-sight connections with ground nodes, rapid deployment, and adjustable mobility \cite{Lit0}, \cite{Lit00}. With such attributes, UAVs are being employed for various applications such as surveillance, rescue missions, delivery, and communication, among others. The development of UAV communication networks has been identified as a promising solution to address the limited coverage area of existing terrestrial communication networks, especially in remote areas.

However, the limited battery life of UAVs is a major concern for their sustainable operation. The conventional UAVs are equipped with communication systems that consume a significant amount of energy, resulting in limited flight time. To overcome this challenge, the concept of reconfigurable intelligent surface (RIS) has emerged as a promising solution to reduce the energy consumption of communication systems. RIS has the potential to enhance the signal-to-interference-plus-noise ratio (SINR) of wireless channels, and thereby reduce the power consumption of the communication system, enabling UAVs to fly for a longer duration. Moreover, RIS can provide communication links to the blocked ground users (GUs) that do not maintain direct access to the UAVs. Therefore, by integrating RISs and UAVs, RIS-assisted UAV networks reduce power consumption of UAVs while improving coverage and connectivity significantly \cite{Lit0}.

In the current literature, communication-efficient  problems were explored for RIS-assisted UAV networks, e.g., \cite{Lit1}-\cite{Lit10}. In particular, the authors of \cite{Lit1} explored an adaptive RIS-assisted aerial-terrestrial downlink communication system between UAVs and multi-users by optimizing RIS element allocation and reflective coefficients. In \cite{Lit2}, the  authors investigated UAV-user communications with RIS assistance to maximize the worst-case secrecy rate, considering transmitter power allocation, RIS beamforming, and UAV trajectory. In \cite{Lit3}, the authors proposed a RIS-assisted UAV communication system to maximize the received signal power at GUs by jointly optimizing passive and active beamforming and UAV trajectory.
Moreover, the work in \cite{Lit4} addressed the energy consumption problem for both orthogonal multiple access (OMA) and non-orthogonal multiple access (NOMA) cases by jointly optimizing the UAV's trajectory and the passive beamforming of the RIS elements. 

Several works also considered the use of RIS-assisted systems \cite{Lit5}, \cite{Lit6}, \cite{Lit7}, which proposed an aerial RIS (ARIS)-assisted system that satisfies the ultra-reliable low latency communication (URLLC) constraints. In \cite{Lit8}, the authors investigated total transmit power minimization for heterogeneous networks that use multiple UAVs and RISs. However, they did not consider energy efficiency which is defined as the ratio of total sum-rate of GUs to the total power consumption of system. Energy efficiency has been widely used to evaluate the performance of conventional communication systems that rely solely on ground based stations \cite{Lit80}, \cite{Lit81}, \cite{Lit82}. Reference \cite{Lit9} focused on maximizing energy efficiency for a single ARIS-assisted downlink communication with a single user and did not consider the multiple ARIS-assisted scenario. In \cite{Lit10}, the authors considered a multiple ARIS configuration to maximize the average energy efficiency for downlink communication between the base station and the GU. The majority of previous works focused on optimizing either the sum-rate or power consumption objective functions. However, most proposed solutions employed computationally expensive methods such as successive convex approximation (SCA) and reinforcement learning. Additionally, there is a higher likelihood of getting trapped in a local minimum when utilizing the SCA method.

Motivated by the aforementioned limitations of the existing related studies, this paper focuses on the energy-efficient communication in RIS-assisted UAV networks. In this context, a RIS is deployed to enhance wireless connectivity while reducing UAV movement. We formulate the problem of maximizing the energy efficiency of the network as a mixed integer non-linear program in which UAV placement, UAV beamforming, On-Off strategy of RIS elements, and phase shift of RIS elements are simultaneously optimized. Since solving this problem is computationally intractable, we use the blocked coordinate descent (BCD) approach. BCD is an optimization algorithm that solves a large-scale optimization problem by iteratively optimizing a subset of variables while holding the other variables fixed. We utilize a continuous genetic algorithm to optimize the phase shift of RIS elements and the beamforming of UAV, whereas a binary genetic algorithm is employed to optimize the On-Off strategy of the RIS elements. Moreover, we optimize the placement of UAV by using the Adam optimizer \cite{adam}. Our simulation results show that our proposed
method offers improved energy efficiency as compared to the existing literature.


\section{System Model and Problem Formulation}\label{S}
\subsection{System Model}
\begin{figure}[t]
\centerline{\includegraphics[height=4.5cm,width=8.5cm]{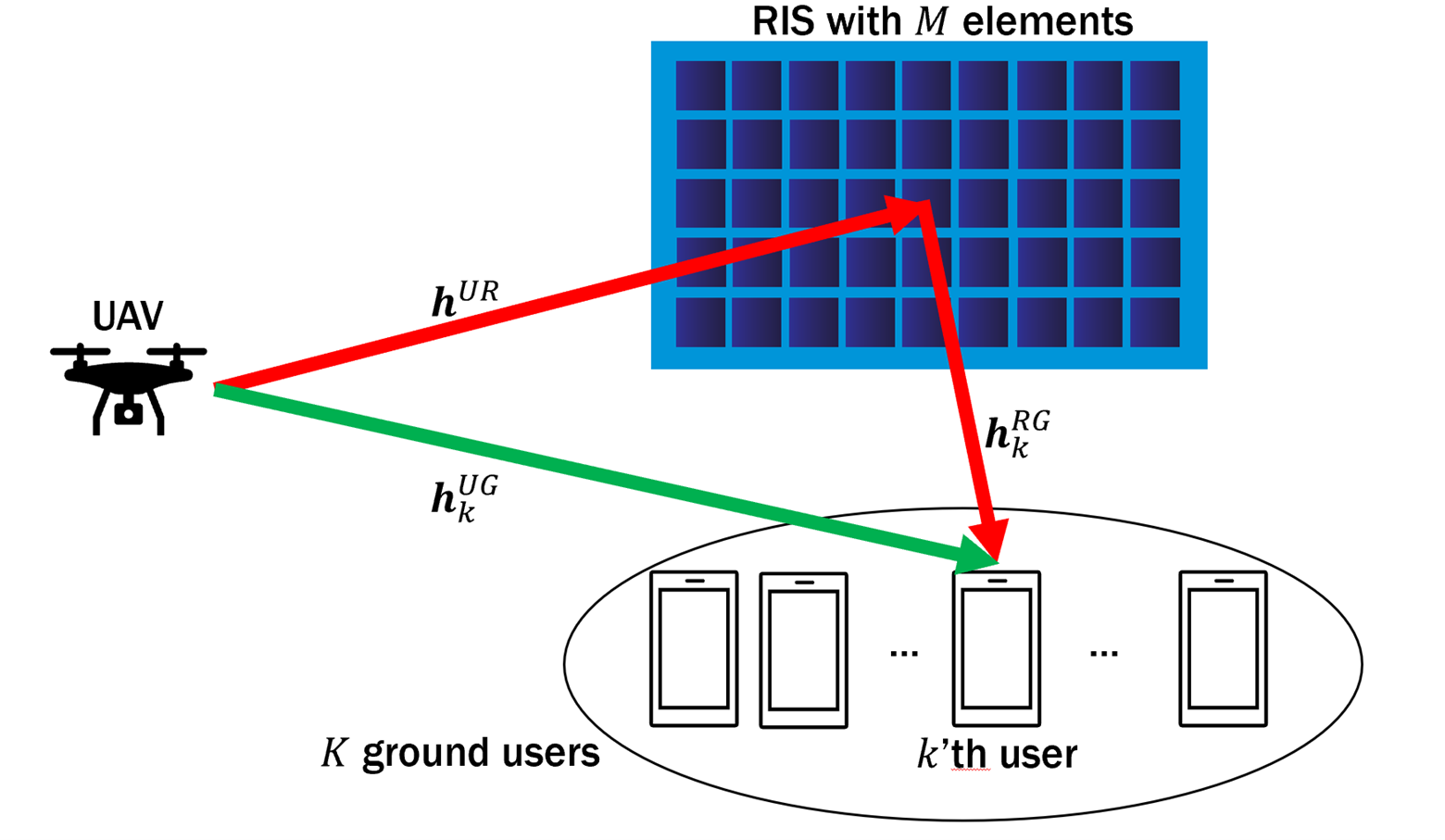}}
\caption{A typical  RIS-assisted UAV communication
system.}
\label{fig11}
\end{figure}

We consider a RIS-assisted UAV network with one UAV, one RIS, and multiple ground users (GUs). All GUs and UAV have single antennas. The UAV can provide good connectivity for
the GUs. To further improve performance, the UAV utilizes a RIS. Fig. \ref{fig11} shows an example of the considered system model. The UAV flies at a fixed altitude $Z_{U}$ for $T$ seconds to provide communication services for $K$ GUs with $0$ altitude. By using a Cartesian coordinate system, we can represent the horizontal position of the UAV, RIS, and the $k$-th GU as $W_{U}=~{[x_{U},y_{U}]}^{T}$, $W_{R}={[x_{R},y_{R}]}^{T}$, and $W_{k}={[x_{k},y_{k}]}^{T}$ for $k=1,2,\ldots,K$. We consider that the RIS and GUs maintain fixed positions, but the position of the UAV can be changed. The RIS is deployed at the altitude of $Z_{R}$ equipped with a controller and $M=M_{c}\times M_{r}$ reflecting units (RUs) to form a uniform array (UA). Specifically, each row of the UA has $M_{r}$ RUs with an equal distance of $d_{r}$ meters and each column of the UA consists of $M_{c}$ RUs with an equal distance of $d_{c}$ meters. These RUs can reflect the received signals in adjustable phase shifts. The phase-shift matrix of the RIS is modeled as the diagonal matrix $\mathbf{\Theta} =diag(e^{j\theta_{1}},e^{j\theta_{2}},\ldots,e^{j\theta_{M}})$, where $\theta_{m}\in [0,2\pi)$ for $m=1,2,\ldots,M$. Considering the power consumption of RISs due to controlling the phase shift values of the reflecting elements \cite{Lit10}, it is often not energy efficient to turn on all the RIS elements. Let us define the binary variable $x_{m}\in \{0,1\}$, where $x_{m}=1$ indicates that the $m$-th RIS element is on, and $0$ otherwise. Defining $\mathbf{\mathcal{X}}=diag(x_{1},\ldots,x_{M})$, the effective phase-shift matrix of the RIS can be written as $\mathbf{\Theta}_{e}=\mathbf{\Theta}\mathbf{\mathcal{X}}$.
\subsection{Channel Model}
The channel between the UAV and the $k$-th GU is given by \cite{Lit102}
\begin{multline*}
\mathbf{h}_{k}^{UG}=\sqrt{\frac{\beta_{0}}{({d_{k}^{UG})}^{\alpha_{k}^{UG}}}}\left(\sqrt{\frac{{\kappa}_{k}^{UG}}{{\kappa}_{k}^{UG}+1}}
+\sqrt{\frac{1}{{\kappa}_{k}^{UG}+1}}\mathbf{\Hat{h}}_{k}^{UG}\right),
\end{multline*}
where $d_{k}^{UG}:=\sqrt{{||W_{k}-W_{U}||}^{2}+Z_{U}^{2}}$ is the distance between the UAV and the $k$-th GU, $\beta_{0}$ is the path loss at the reference distance of $1$ meter, $\alpha_{k}^{UG}$ represents the path loss exponent of the link from the UAV to the $k$-th GU, ${\kappa}_{k}^{UG}$ denotes the corresponding Rician factor, and $\mathbf{\Hat{h}}_{k}^{UG}\sim \mathcal{CN}(\mathbf{0},1)$ is the scattering component of the $k$-th GU.

The link from the UAV to the RIS is presumed by a line-of-sight (LOS) channel, and the channel gain can be found by \cite{Lit103}
\begin{equation}\nonumber
\mathbf{h}^{UR}=\sqrt{\frac{\beta_{0}}{({d_{}^{UR})}^{2}}}\mathbf{h}_{LOS}^{UR},
\end{equation}
where $d_{}^{UR}:=\sqrt{{||W_{R}-W_{U}||}^{2}+{\left(Z_{R}-Z_{U}\right)}^{2}}$ is the distance between the UAV and the RIS, and $\mathbf{h}_{LOS}^{UR}$ represents the array response component which can be denoted by
\begin{eqnarray}\nonumber
&~&\mathbf{h}_{LOS}^{UR}=\\\nonumber
&~&{[1,e^{-j\frac{2\pi d_{r}}{\lambda}\phi^{UR}\psi^{UR} },\ldots,e^{-j\frac{2\pi d_{r}}{\lambda}(M_{r}-1)\phi^{UR}\psi^{UR} }]}^{T}\\\nonumber
&~&\otimes {[1,e^{-j\frac{2\pi d_{c}}{\lambda}\varphi^{UR}\psi^{UR} },\ldots,e^{-j\frac{2\pi d_{c}}{\lambda}(M_{c}-1)\varphi^{UR}\psi^{UR} }]}^{T}
\end{eqnarray}
where $\phi^{UR}=\frac{y_{U}-y_{R}}{||W_{R}-W_{U}||}$, $\varphi^{UR}=\frac{x_{R}-x_{U}}{||W_{R}-W_{U}||}$, and $\psi^{UR}=~\frac{z_{U}-z_{R}}{d_{}^{UR}}$. We point out that $\phi^{UR}$, $\varphi^{UR}$, and $\psi^{UR}$ are the cosine of horizontal angle-of-arrival (AoA), sine of horizontal AoA, and sine of vertical AoA, respectively.

Considering the channel between RIS and each GU, we assume that the RIS-aided channels are modeled by the Rician fading. Thus, the channel gain is given by \cite{Lit102}
\begin{multline*}
\mathbf{h}_{k}^{RG}=\sqrt{\frac{\beta_{0}}{({d_{k}^{RG})}^{\alpha_{k}^{RG}}}}\bigl(\sqrt{\frac{{\kappa}_{k}^{RG}}{{\kappa}_{k}^{RG}+1}}\mathbf{h}_{k,LOS}^{RG}\\
+\sqrt{\frac{1}{{\kappa}_{k}^{RG}+1}}\mathbf{\Hat{h}}_{k}^{RG}\bigr),
\end{multline*}
where $d_{k}^{RG}:=\sqrt{{||W_{k}-W_{R}||}^{2}+Z_{R}^{2}}$ is the distance between the RIS and the $k$-th GU, $\alpha_{k}^{RG}$ denotes the path loss exponent of the link between the RIS and the $k$-th GU, ${\kappa}_{k}^{RG}$ is the corresponding Rician factor, $\mathbf{\Hat{h}}_{k}^{RG}\sim \mathcal{CN}(\mathbf{0},I_{M})$ is the scattering component, and the LOS component is
\begin{eqnarray}\nonumber
&~&\mathbf{h}_{k,LOS}^{RG}=\\\nonumber
&~&{[1,e^{-j\frac{2\pi d_{r}}{\lambda}\phi_{k}^{RG}\psi_{k}^{RG} },\ldots,e^{-j\frac{2\pi d_{r}}{\lambda}(M_{r}-1)\phi_{k}^{RG}\psi_{k}^{RG} }]}^{T}\\\nonumber
&~&\otimes {[1,e^{-j\frac{2\pi d_{c}}{\lambda}\varphi_{k}^{RG}\psi_{k}^{RG} },\ldots,e^{-j\frac{2\pi d_{c}}{\lambda}(M_{c}-1)\varphi_{k}^{RG}\psi_{k}^{RG} }]}^{T},
\end{eqnarray}
in which $\phi_{k}^{RG}=\frac{y_{k}-y_{R}}{||W_{k}-W_{R}||}$, $\varphi_{k}^{RG}=\frac{x_{k}-x_{R}}{||W_{k}-W_{R}||}$, and $\psi_{k}^{RG}=~\frac{z_{R}}{d_{k}^{RG}}$. In a similar way, $\phi_{k}^{RG}$, $\varphi_{k}^{RG}$, and $\psi_{k}^{RG}$ denote cosine of horizontal angle-of-departure (AoD), sine of horizontal AoD, and sine of vertical AoD, respectively.

Given the aforementioned channel models, the effective channel gain between the UAV and the $k$-th GU with the aid of the RIS is given by $\mathbf{C}_{k}=\mathbf{h}_{k}^{UG}+{(\mathbf{h}_{k}^{RG})}^{H}\mathbf{\Theta}_{e}\mathbf{h}^{UR}$.

\subsection{Problem Formulation}
In this paper, the UAV is assumed to share the same frequency band for providing services to the GUs. The transmitted signal of the UAV is given by $\mathbf{S}=\sum\limits_{k=1}^{K}\sqrt{p_{k}}s_{k}$, where $p_{k}$ and $s_{k}$ are the transmitted power and signal for the $k$-th GU, respectively. Assuming that $P_{max}$ is the maximum transmit power of the UAV, we have $\sum\limits_{k=1}^{K}p_{k}\leq P_{max}$. Then the received signal at the $k$-th GU can be expressed as
\begin{equation}\nonumber
y_{k}=\mathbf{C}_{k}\sqrt{p_{k}}s_{k}+\underbrace{\mathbf{C}_{k}\sum\limits_{t=1,t\neq k}^{K}\sqrt{p_{t}}s_{t}}_{\text{interference}}+\underbrace{n_{k}}_{\text{noise}},
\end{equation}
where $n_{k}\sim \mathcal{CN}(0,\sigma^{2})$.
Therefore, the received signal-to-interference-plus-noise ratio (SINR) at the $k$-th GU is 
\begin{equation}\nonumber
\gamma_{k}=\frac{{|{\mathbf{C}}_{k}|}^{2}p_{k}}{{|{\mathbf{C}}_{k}|}^{2}\sum\limits_{t=1,t\neq k}^{K}p_{t}+\sigma^{2}}.
\end{equation}
As a result, the sum-rate of all GUs can be written as
\begin{equation}
R_{t}=B\sum\limits_{k=1}^{K}\log_{2}(1+\gamma_{k}),
\end{equation}
where $B$ is the bandwidth of the channel.

Essentially, the UAV consumes power due to hovering and wireless signal transmission, while GUs consume power for circuit operations and the RIS elements consume power for controlling the phases. 
Consequently, the total power consumption of our envisioned system is given by
\begin{multline}\label{totpow}
P_{t}=\underbrace{p_{h}}_{\text{hovering power of UAV}}+\underbrace{\sum\limits_{k=1}^{K}p_{k}}_{\text{transmit power of UAV}}\\
+\underbrace{\sum\limits_{k=1}^{K}p_{k}^{c}}_{\text{circuit power of GUs}}+\underbrace{\sum\limits_{m=1}^{M}p^{r}x_{m}}_{\text{power consumption of RIS}},
\end{multline}
where $p_{k}^{c}$ is the circuit power consumption of the $k$-th GU and $p^{r}$ is the power consumption of each RU. In (\ref{totpow}), $p_{h}$ is the drone's hovering power which is expressed by $p_{h}=\sqrt{\frac{{{(mg)}}^{3}}{2\pi r_{p}^{2}n_{p}\rho}}$, where $m$ is the drone’s weight, $g$ is the gravitational acceleration of the earth, $r_{p}$ is propellers’ radius, $np$ is the number of propellers, and $\rho$ is the air density \cite{Lit11}. From (1) and (2), we can define the energy efficiency of the RIS-assisted UAV network as $\eta=\frac{R_{t}}{P_{t}}$.

Given the considered system model, our objective is to jointly optimize the phase-shift of RIS elements, RIS On-Off vector, beamforming vector of the UAV, and its placement to maximize the energy efficiency under the minimum rate requirement of GUs and the total power constraint of the UAV. Mathematically, the energy  efficiency maximization optimization problem can be formulated as
\begin{equation*}
\begin{aligned}
& \underset{\mathbf{X},\boldsymbol{\theta},\mathbf{P},W_{U}}{\text{maximize}}
& & \eta \\
& \text{subject to}
& &B\log_{2}(1+\gamma_{k})\geq R_{k,min},~~~~~~~~~~(3a)\\
&&& \sum\limits_{k=1}^{K}p_{k}\leq P_{max},~~~~~~~~~~(3b)
\\
&&& p_{k}>0,~~~~~\text{for}~1 \leq k \leq K,~~~~~~~~~~(3c)
\\
&&& x_{m}\in \{0,1\},~~~~~\text{for}~1 \leq m \leq M
\\
&&& \theta_{m}\in [0,2\pi),~~~~~\text{for}~1 \leq m \leq M
\end{aligned}
\label{opsti}
\end{equation*}
where $\mathbf{X}={(x_{1},\ldots,x_{M})}^{T}$, $\boldsymbol{\theta}={(\theta_{1},\ldots,\theta_{M})}^{T}$, $\boldsymbol{P}=~{(p_{1},\ldots,p_{K})}^{T}$, and $R_{k,min}$ is the minimum data rate requirement of the $k$-th GU. The minimum rate constraint for each GU is given in (3a). Moreover, (3b) and (3c) represent the total power constraint of the UAV.

The derived optimization problem is a mixed-integer nonlinear program (MINLP) even for the single GU case with $K = 1$. It is generally difficult to obtain the globally optimal solution of a MINLP. 

\section{Proposed Solution}\label{PS}
In this section, we first use the alternating optimization approach (BCD) to divide the derived MINLP into three sub-problems, which has been widely applied to tackle the non-concave problems in
RIS-assisted UAV networks \cite{Lit103}. Then each sub-problem is separately optimized, while other variables are considered to be fixed. Specifically, we employ a continuous genetic algorithm to jointly optimize the phase shifts of the RIS and beamforming of UAV, and a binary genetic algorithm to optimize the On-Off strategy of the RIS. Afterwards, the optimized position of the UAV is obtained using the Adam optimizer. The block diagram of the overall proposed solution is depicted in Fig. 2. As it can be observed, the optimization process terminates if the amount of improvement in energy efficiency is less than a given threshold $\delta$.
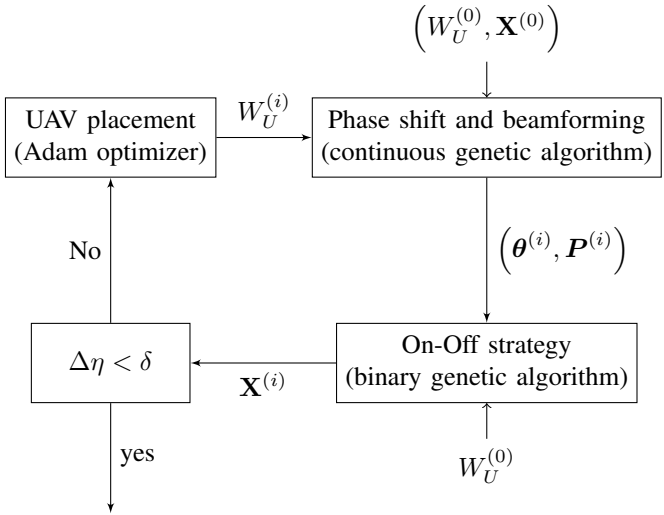
\begin{figure}\label{blocki}
\tikzstyle{block} = [draw, fill=white, rectangle, 
    minimum height=3em, minimum width=6em]
\tikzstyle{input} = [coordinate]
\tikzstyle{output} = [coordinate]
\tikzstyle{pinstyle} = [pin edge={to-,thin,black}]

\begin{tikzpicture}[auto, node distance=2cm,>=latex']

    \node [block, align=center] (controller) {UAV placement \\ (Adam optimizer)};
    
    \node [block, right of=controller, pin={[pinstyle]above:$\left(W_{U}^{(0)},{\mathbf{X}}^{(0)}\right)$},
            node distance=5cm, align=center] (system) {Phase shift and beamforming \\ (continuous genetic algorithm)};
    \node [block, below of=system, pin={[pinstyle]below:$W_{U}^{(0)}$},
            node distance=3cm, align=center] (onoff) {On-Off strategy \\ (binary genetic algorithm)};
    
    \node [block, below of=controller, node distance=3cm] (measurements) {$\Delta \eta < \delta$};
    \node [output, below of=measurements] (output) {};
    \draw [draw,->] (measurements) -- node {No} (controller);
    \draw [draw,->] (system) -- node {$\left(\boldsymbol{\theta}^{(i)},\boldsymbol{P}^{(i)}\right)$} (onoff);
    \draw [draw,->] (controller) -- node {$W_{U}^{(i)}$} (system);
    \draw [draw,->] (onoff) -- node {${\mathbf{X}}^{(i)}$} (measurements);
    \draw [->] (measurements) -- node [name=y] {yes}(output);
        

\end{tikzpicture}
\caption{Block diagram of the overall proposed solution.}
\end{figure}

\begin{algorithm}[t]
 \begin{algorithmic}
 \renewcommand{\algorithmicrequire}{\textbf{Input: Fitness function $\eta(\boldsymbol{\theta},\boldsymbol{P})$ and population size $2L$}}
 \STATE Generate the initial population $\mathbb{P}^{(0)}= {[\mathbf{Y}_{1}^{(0)},\mathbf{Y}_{2}^{(0)},\ldots,\mathbf{Y}_{2L}^{(0)}]}^{T}$.
 \STATE Let $Q^{(0)}$ be a selection random variable defined on the set $\{1,2,\ldots,2L\}$ with probability mass function $Pr\{Q^{(0)}=k\}=\frac{\eta\left({\mathbf{Y}_{k}^{(0)}}\right)}{\sum\limits_{l=1}^{2L}\eta\left({\mathbf{Y}_{l}^{(0)}}\right)}$.
 \FOR {$i=1:N_{1}$}
 \STATE Select $2L$ individuals from $\mathbb{P}^{(i-1)}$ using the selection random variable $Q^{(i-1)}$.
 \STATE Randomly select $L$ pairs of individuals in $\mathbb{P}^{(i-1)}$ for cross-over. Substitute each pair with the two generated offsprings.
 \STATE Add a small random value to each element of
 generated offsprings (mutation).
 \STATE Find the updated population $\mathbb{P}^{(i)}$ and selection random variable $Q^{(i)}$.
 \ENDFOR
 \end{algorithmic} 
 \caption{Continuous genetic algorithm for phase shift and beamforming optimization}
 \end{algorithm}

\begin{algorithm}[t]
\caption{Adam optimizer for the UAV placement. $g_{i}^{2}$ indicates the element-wise square $g_{i}\odot g_{i}$. Good default settings are $\alpha=0.001$, $\beta_{1}=0.9$, $\beta_{2}=0.999$, and $\epsilon=10^{-8}$. All operations on vectors are element-wise. With $\beta_{1}^{i}$ and $\beta_{2}^{i}$, we denote $\beta_{1}$ and $\beta_{2}$ to the power of $i$.}
\label{alg:adam}
\begin{algorithmic}
\REQUIRE $\alpha$: Step size
\REQUIRE $\epsilon$: Small constant
\REQUIRE $\beta_{1},\beta_{2} \in [0,1)$: Exponential decay rates for the moment estimates
\REQUIRE $\eta\left(W_{U}\right)$: Objective function 
\REQUIRE $W_{U}^{(0)}$: Initial position of UAV
\STATE Initialize $m_0 \leftarrow 0$, $v_0  \leftarrow  0$, and $i  \leftarrow  0$
\FOR {$i=1:N_{2}$}
    \STATE Compute gradient $g_i \gets \nabla_{W_{U}}\eta\left(W_{U}^{(i-1)}\right)$.
    \STATE Update biased first moment estimate: $m_i \leftarrow \beta_1 m_{i-1} + (1-\beta_1) g_i$
    \STATE Update biased second raw moment estimate: $v_i \leftarrow \beta_2 v_{i-1} + (1-\beta_2) g_{i}^{2}$
    \STATE Compute bias-corrected first moment estimate: $\hat{m}_i \leftarrow \frac{m_i}{1-\beta_1^i}$
    \STATE Compute bias-corrected second raw moment estimate: $\hat{v}_i \leftarrow \frac{v_i}{1-\beta_{2}^{i}}$
    \STATE Update parameters: $W_{U}^{(i)} \gets W_{U}^{(i-1)} - \alpha \frac{\hat{m}_i}{\sqrt{\hat{v}_i} + \epsilon}$
\ENDFOR
\end{algorithmic}
\end{algorithm}

\subsection{Phase Shift and Beamforming Optimization}
For optimizing the phase shift of the RIS elements and beamforming of the UAV, we take advantage of the continuous genetic algorithm. The continuous genetic algorithm is a type of optimization algorithm that is used to search for optimal solutions in continuous parameter spaces \cite{Lit13}. The algorithm is based on the principles of natural selection and evolution, and it works by creating a population of candidate solutions and iteratively refining them over multiple generations. The main steps of the continuous genetic algorithm include initialization, selection, crossover, mutation, and termination.
\begin{itemize}
\item The first step of the algorithm is initialization, where a population of potential solutions is created randomly. The size of the population is typically set based on the complexity of the problem, and each candidate solution is represented as a vector of continuous values.
\item In the selection step, the candidate solutions are evaluated based on a fitness function, which measures how well they perform on the given problem. In this paper, energy efficiency ($\eta$) is considered as the fitness function. The best-performing solutions are selected to become parents for the next generation. 
\item In the crossover step, pairs of parents are combined to produce new offspring solutions. This is done by finding the weighted sum of parent vectors and creating a new solution vector that inherits traits from both parents. The crossover process helps to explore different regions of the search space and can help to create more diverse candidate solutions.
\item In the mutation step, random changes are introduced into the offspring solutions to further explore the search space and prevent the algorithm from getting stuck in a local optimum. This is done by adding a small random value to each element of the solution vector.
\end{itemize}
The algorithm continues to iterate through the selection, crossover, and mutation steps until a termination criterion is met. The details of the continuous genetic algorithm for optimizing the phase shift of the RIS elements and beamforming of the UAV are given in Algorithm 1.

\subsection{RIS On-Off Optimization}
Given UAV position $W_{U}$, phase vector $\boldsymbol{\theta}$, and beamforming vector $\boldsymbol{P}$, problem (3) is a nonlinear integer optimization problem with respect
to the RIS On-Off vector $\mathbf{X}$. Since the nonlinear integer optimization problem is NP-hard in general, it is difficult to obtain the globally optimal solution with polynomial complexity. To tackle this computational intractability, we use the binary genetic algorithm to solve the RIS On-Off optimization problem. The main steps of binary genetic algorithm are same as the continuous one, where crossover and mutation steps are adapted to be applied to the vectors of binary values. In the crossover step,  parts of the parent vectors are swapped to create a new solution vector, whereas binary bits of offspring solutions are flipped with a small probability in the mutation step.

\subsection{UAV Placement Optimization}
To obtain the optimized position of the UAV, we utilize the Adam optimizer. The Adam optimizer is a gradient-based optimization algorithm that is widely used in machine learning because of its superior performance compared to other gradient descent methods \cite{adam}. One of the primary reasons for its superiority is that it maintains separate learning rates for each parameter and adapts them based on the historical gradient information. This allows the algorithm to handle sparse gradients and noisy data more efficiently, leading to faster convergence and better performance. Additionally, the adaptive learning rates ensure that the algorithm can navigate complex loss landscapes more effectively, avoiding getting stuck in local minimum.

The Adam optimizer works by computing a moving average of the gradient and its square, which allows it to adjust the learning rate adaptively. The algorithm consists of several steps, including initializing the moving averages, computing the gradient on a mini-batch of data, updating the moving averages with the gradient information, and finally updating the parameters based on the adaptive learning rate. The algorithm also incorporates bias correction terms to ensure that the moving averages are initialized at zero and that the updates are unbiased. The detailed steps of the Adam optimizer for optimizing the UAV placement are given in Algorithm 2. 

\begin{figure}[t]
\centerline{\includegraphics[height=5cm,width=8.5cm]{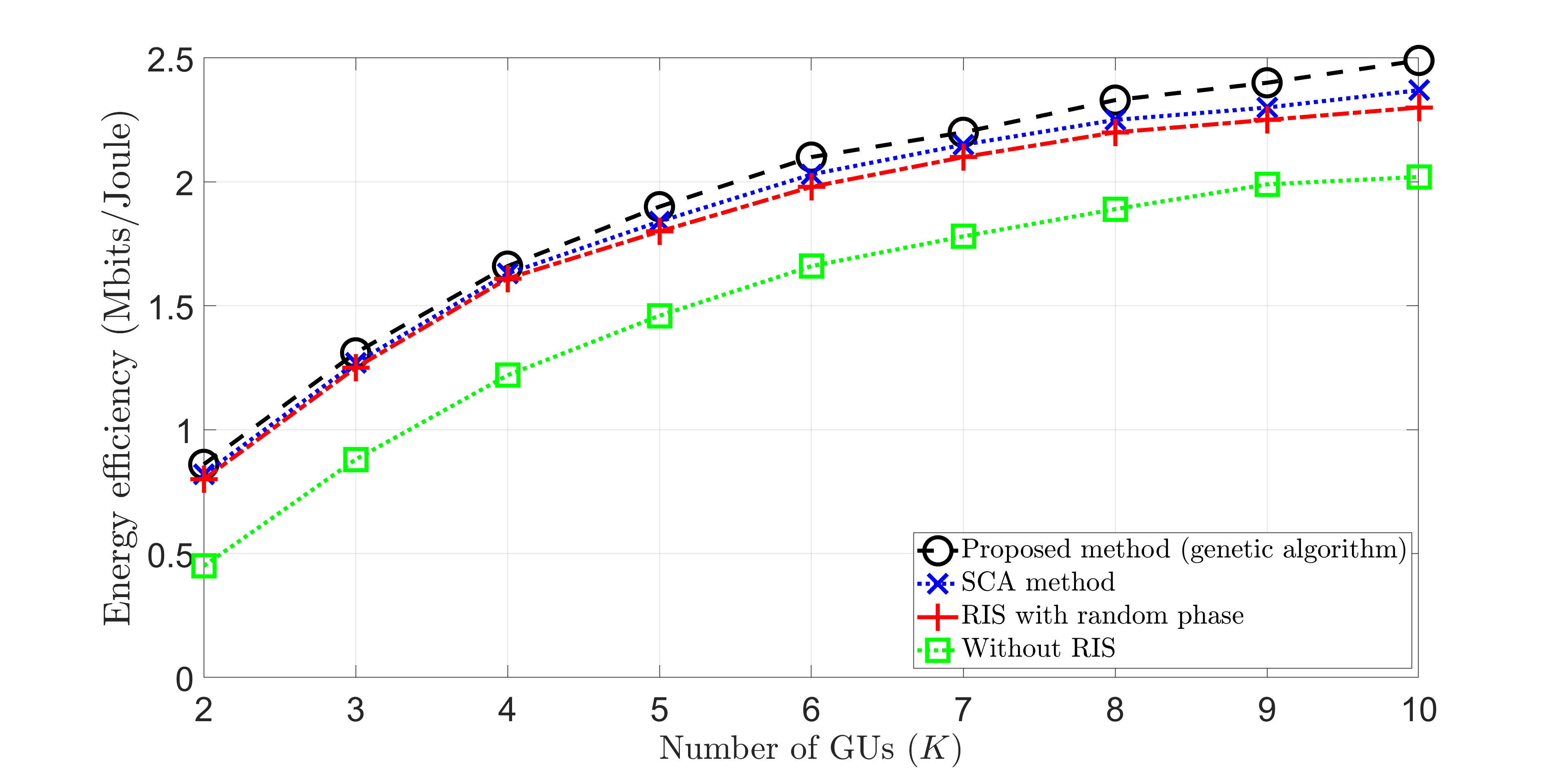}}
\caption{Energy efficiency versus the number of GUs when $M=60$.}
\label{figgu1}
\end{figure}

\section{Numerical Results}\label{NR}
This section presents numerical results that validate the efficacy of the proposed alternating approach for optimizing UAV placement and beamforming, On-Off strategy of RIS elements, and phase shift of RIS elements.  The following benchmark schemes are used for performance comparison:
\begin{itemize}
\item \textbf{Without RIS}: This scheme does not employ the RIS to assist communication and considers only the direct link between the UAV and the GUs.
\item \textbf{RIS with random Phase}: In this scheme, the phase shift of the RIS elements are chosen randomly within the range of $[0,2\pi)$.
\item \textbf{SCA method}: In this scheme, we apply SCA method to optimize the phase shift of the RIS elements. SCA is an iterative optimization technique used to solve non-convex optimization problems \cite{Lit8}. It involves transforming a non-convex optimization problem into a sequence of convex sub-problems, which can be solved efficiently.
\end{itemize}

In this paper, we examine a scenario similar to the configuration presented in \cite{Lit103}, where a single RIS is installed on a building facade to improve aerial-ground communications. The GUs are distributed randomly and uniformly within a circular area centered at $(200~m, 25~m)$ with a radius of $20$ m. The details of the considered simulation parameters are given in Table I.

Assuming that there are 60 RIS elements, Fig. \ref{figgu1} illustrates the relationship between the energy efficiency and the number of GUs. The figure shows that the energy efficiency increases as the number of GUs ($K$) increases. Additionally, it is apparent that not using RIS results in consistently low energy efficiency since there is no indirect path from the UAV to the GUs. Overall, our proposed method is superior to the SCA method and random phase, particularly for a large number of GUs since the SCA method is more prone to trapping in a local minimum.

\begin{figure}[t]
\centerline{\includegraphics[height=5cm,width=8.5cm]{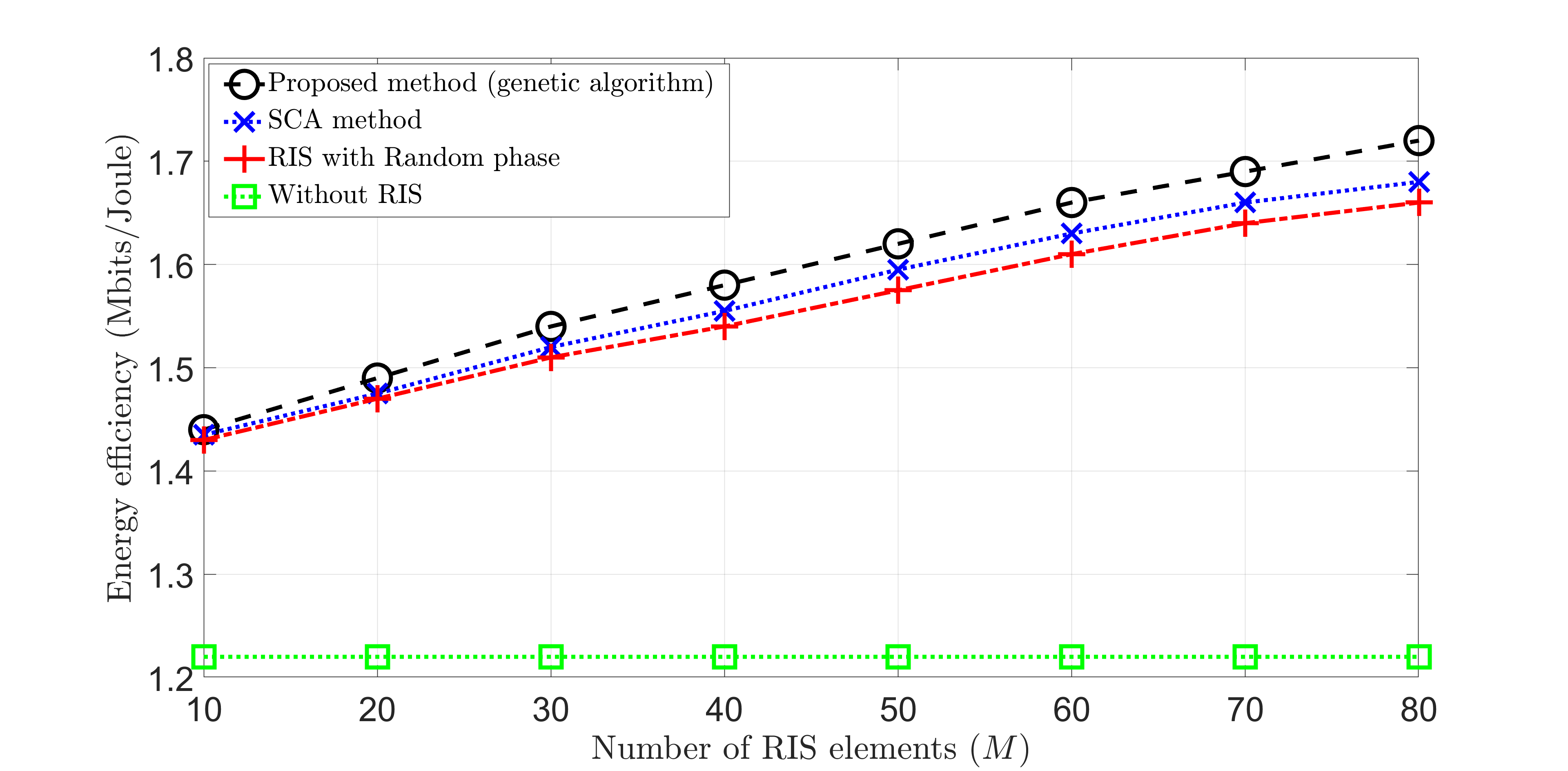}}
\caption{Energy efficiency for different number of RIS elements when $K=4$.}
\label{figguurr2}
\end{figure}

Fig. \ref{figguurr2} illustrates the relationship between the energy efficiency  and the number of the RIS elements ($M$) when $K = 4$. Fig. \ref{figguurr2} clearly shows that deploying RIS significantly improves the system performance, whereas negligible gains are observed without the RIS. Optimizing phase shifts results in slightly better performance compared to the random phase shift scheme, emphasizing the importance of optimizing phase shifts. The proposed method outperforms the SCA method, and this improvement becomes more apparent as the number of GUs increases.

In summary, integrating RIS with UAV communications can significantly enhance network coverage, and the energy efficiency of the RIS-assisted schemes improves as the number of RIS elements increases. This suggests that a larger number of reflecting elements provides more flexibility for the joint phase shift and UAV placement optimization, leading to better communication quality and higher gains.

\begin{table}[t]
\centering
\caption{Simulation parameters}
\begin{tabular}{|c|c|c|c|}
\hline
parameter & value & parameter & value \\ 
\hline
\hline
$Z_{U}$ & 70 m & $Z_{R}$ & 40 m \\ 
\hline 
$W_{U}^{(0)}$ & (200 m, 50 m) & $W_{R}$ & (200 m, 0 m) \\
\hline
$\lambda$ & 10 cm & $P_{max}$ & 1 W \\
\hline
 $d_{r}$ & 5 cm  & $d_{c}$ & 5 cm \\
\hline
 $B$  &  20 MHz & $\beta_{0}$  & $10^{-2}$  \\
\hline
$\alpha_{k}^{UG}$ & 3  & $\alpha_{k}^{RG}$  & 2.4\\
\hline
$\kappa_{k}^{UG}$ & 2   & $\kappa_{k}^{RG}$ & 2 \\
\hline
$p_{k}^{c}$ & 1 mW  & $p^{r}$  &  1 mW\\
\hline
 $m$ &  2 kg & $g$ & 9.8 $\frac{m}{s^{2}}$ \\
\hline
$r_{p}$ &  0.2  &  $n_{p}$  & 4 \\
\hline
 $\rho$ &  1.225 $\frac{kg}{m^{3}}$   & $R_{k,min}$ & 100 $\frac{bits}{s}$ \\
 \hline
\end{tabular}
\end{table}


\section{Conclusion}\label{C}
In this paper, we proposed an energy-efficient communication scheme for RIS-assisted UAV networks. By optimizing the On-Off strategy, phase shift of RIS elements, and the UAV placement and beamforming, our proposed scheme maximized the energy efficiency of the network. We formulated the optimization problem as a mixed integer nonlinear program and utilized the BCD approach to tackle it. We employed a combination of continuous and binary genetic algorithms, as well as the Adam optimizer, to optimize the various variables of the problem. Our simulation results showed that the proposed scheme outperforms existing literature in terms of energy efficiency. Specifically, our scheme achieves significant improvements in the energy efficiency of downlink communications between the aerial base station and the GUs.



\end{document}